\documentclass[12pt,a4paper,final]{iopart}

\usepackage{graphicx}
\usepackage[breaklinks=true,colorlinks=true,linkcolor=blue,urlcolor=blue,citecolor=blue]{hyperref}
\usepackage{iopams}
\expandafter\let\csname equation*\endcsname\relax

\expandafter\let\csname endequation*\endcsname\relax

\usepackage{amsmath}

\begin{document}

\title[Analytic Response Relativistic Coupled-Cluster Theory]{Analytic Response Relativistic Coupled-Cluster Theory: The first application to indium isotope shifts}

\author{B.K.~Sahoo}
\address{Atomic and Molecular Physics Division, Physical Research Laboratory, Navrangpura, Ahmedabad 380009, India}
\ead{bijaya@prl.res.in}
\author[cor1]{A.R.~Vernon}
\address{School of Physics and Astronomy, The University of Manchester, Manchester M13 9PL, United Kingdom}
\ead{\mailto{adam.vernon@cern.ch}, \mailto{adam.vernon-2@postgrad.manchester.ac.uk}}
\author{R.F. Garcia Ruiz}
\address{School of Physics and Astronomy, The University of Manchester, Manchester M13 9PL, United Kingdom}
\address{EP Department, CERN, CH-1211 Geneva 23, Switzerland}
\ead{ronald.fernando.garcia.ruiz@cern.ch}
\author{C.L.~Binnersley}
\address{School of Physics and Astronomy, The University of Manchester, Manchester M13 9PL, United Kingdom}
\author{J.~Billowes}
\address{School of Physics and Astronomy, The University of Manchester, Manchester M13 9PL, United Kingdom}
\author{M.L.~Bissell}
\address{School of Physics and Astronomy, The University of Manchester, Manchester M13 9PL, United Kingdom}
\author{T.E.~Cocolios}
\address{KU Leuven, Instituut voor Kern- en Stralingsfysica, B-3001 Leuven, Belgium}
\author{G.J.~Farooq-Smith}
\address{KU Leuven, Instituut voor Kern- en Stralingsfysica, B-3001 Leuven, Belgium}
\author{K.T.~Flanagan}
\address{School of Physics and Astronomy, The University of Manchester, Manchester M13 9PL, United Kingdom}
\address{Photon Science Institute Alan Turing Building, University of Manchester, Manchester M13 9PY, United Kingdom}
\author{W.~Gins}
\address{KU Leuven, Instituut voor Kern- en Stralingsfysica, B-3001 Leuven, Belgium}
\author{R.P.~de~Groote}
\address{KU Leuven, Instituut voor Kern- en Stralingsfysica, B-3001 Leuven, Belgium}
\address{Department of Physics, University of Jyv\"askyl\"a, Survontie 9, Jyv\"askyl\"a, FI-40014, Finland}
\author{\'A.~Koszor\'us}
\address{KU Leuven, Instituut voor Kern- en Stralingsfysica, B-3001 Leuven, Belgium}
\author{G.~Neyens}
\address{EP Department, CERN, CH-1211 Geneva 23, Switzerland}
\address{KU Leuven, Instituut voor Kern- en Stralingsfysica, B-3001 Leuven, Belgium}
\author{K.M.~Lynch}
\address{EP Department, CERN, CH-1211 Geneva 23, Switzerland}
\author{F. Parnefjord-Gustafsson}
\address{KU Leuven, Instituut voor Kern- en Stralingsfysica, B-3001 Leuven, Belgium}
\author{C.M.~Ricketts}
\address{School of Physics and Astronomy, The University of Manchester, Manchester M13 9PL, United Kingdom}
\author{K.D.A~Wendt}
\address{Institut f\"ur Physik, Johannes Gutenberg-Universit\"at Mainz, D-55128 Mainz, Germany}
\author{S.G.~Wilkins}
\address{EN Department, CERN, CH-1211 Geneva 23, Switzerland}
\author{X.F.~Yang}
\address{School of Physics and State Key Laboratory of Nuclear Physics and Technology, Peking University, Beijing 100871, China}
\address{KU Leuven, Instituut voor Kern- en Stralingsfysica, B-3001 Leuven, Belgium}

\begin{abstract}
With increasing demand for accurate calculation of isotope shifts of atomic systems for fundamental and nuclear structure research, an analytic energy derivative approach is presented in the relativistic coupled-cluster theory framework to determine the atomic field shift and mass shift factors. This approach allows the determination of expectation values of atomic operators, overcoming fundamental problems that are present in existing atomic physics methods, i.e. it satisfies the Hellmann-Feynman theorem, does not involve any non-terminating series, and is free from choice of any perturbative parameter. As a proof of concept, the developed analytic response relativistic coupled-cluster theory has been applied to determine mass shift and field shift factors for different atomic states of indium. High-precision isotope-shift measurements of $^{104-127}$In were performed in the 246.8-nm (5p~$^2$P$_{3/2}$ $\rightarrow$ 9s~$^2$S$_{1/2}$) and 246.0-nm (5p~$^2$P$_{1/2}$ $\rightarrow$ 8s~$^2$S$_{1/2}$) transitions to test our theoretical results. An excellent agreement between the theoretical and measured values is found, which is known to be challenging in multi-electron atoms. The calculated atomic factors allowed an accurate determination of the nuclear charge radii of the ground and isomeric states of the $^{104-127}$In isotopes, providing an isotone-independent comparison of the absolute charge radii.
\end{abstract}

\pacs{00.00, 20.00, 42.10}
\vspace{2pc}
\noindent{\it Keywords}: Article preparation, IOP journals
\submitto{\NJP}

\section{Introduction}
The removal or addition of neutrons to the nucleus produces changes in the energy of atomic transitions, known as the isotope shift (IS). These small variations, typically less than 10$^{-6}$ with respect to the atomic energy levels, can probe fundamental aspects of the electron-nucleus interaction, e.g., the size of the nucleus \cite{garciaruiz16}, the existence of new bosons \cite{berengut18,jen18}, new spin-independent interactions \cite{dela17a,dela17b} and long-range neutrino-mediated forces  \cite{stadnik18}. Currently, extensive experimental efforts worldwide have been focused on the development of complementary techniques to perform high-precision measurements of IS in atomic transitions, across different isotopic chains \cite{Marsh2018,Gebert15,Hammen18,raeder18}. Alongside the experimental progress, the development of many-body methods plays a central role in these studies as it provides the means to extract nuclear-structure and fundamental-physics parameters from experimental observations \cite{Safronova2018}. Reliable atomic calculations are critical to establish firm conclusions from high-precision experiments in nuclear \cite{Cheal2012} and fundamental-physics research \cite{Flambaum18}.

Most of our present knowledge on the nuclear charge radius of unstable nuclei is derived from IS measurements in atomic transitions performed by laser spectroscopy experiments \cite{Cheal2012}. The calculation of atomic physics factors which are needed to decouple many-body electron correlations from nuclear-structure variations present the main challenges in the interpretation of IS measurements. The coupled-cluster (CC) method is considered as the gold standard for treating electron-correlation effects \cite{Bishop1991}. However, the current methods used to calculate atomic physics operators present serious drawbacks that can generate uncontrolled theoretical uncertainties. The commonly used expectation-value-evaluation (EVE) approach \cite{Sahoo2018,Sahoo2019}, for example, involves non-terminating series, and the finite-field (FF) approach \cite{Vasiliev1997} depends on the choice of a perturbation parameter. To overcome these problems in this work we implement and demonstrate, for the first time in atomic systems, the analytic-response (AR) theory within the CC framework \cite{Monkhorst1977} to determine IS shift parameters of atomic systems.

The atomic factors involved in the IS measurements can be empirically obtained for even-proton elements \cite{Frickeb}, where independent charge radii measurements from electron scattering and muonic atoms exists for three or more stable isotopes. However, this is not the case for elements with odd-proton number, where only up to two stable isotopes exists and the accuracy of all charge-radii values obtained from isotope shifts measurements relies on atomic physics calculations.
Accurate determination of the charge radii of radioactive isotopes is not only relevant for nuclear structure research, but can provide a deeper insight into nuclear matter \cite{Brown17,yang18}. Motivated by the current nuclear structure interest in the study of ISs around proton number $Z$~=~50 \cite{Gorges2019,Hammen2018,rej16,gar17}, our theoretical developments were used to perform, for the first time, {\it ab-initio} calculations of atomic factors for indium (In) atom ($Z$~=~49). The In isotope chain offers a comprehensive laboratory to test these theoretical developments.
The long chain of isotopes increases the precision in cancelling out the nuclear contribution to the IS, while the presence of at least one isomeric nuclear state at each mass allows for an mass-independent measure of the field-shift (FS) contribution to the IS. This provides a stringent constraint to test our theoretical calculations by increasing the precision on the experimentally determined atomic factors. Moreover, several atomic transitions are experimentally accessible, and precise data on transitions to high-lying states \cite{Menges1985} can be combined with our new measurements and calculations to evaluate the individual atomic level-IS (LIS), allowing a direct study of the IS factors for each level.

\section{Theory}

The IS of an energy level, $i$, between an isotope, $A$, with mass, $m_A$, and an isotope, $A'$, with mass, $m_{A'}$, is given \cite{Foot2004} by a product of nuclear and atomic factors as \footnote{The factor of $h$ is dropped in the notation of this work unless relevant i.e. IS$=\delta E_i$. However where values are presented for comparison to experiment the factor is included.}
\begin{eqnarray}
\delta E_i &=& F_i \delta \langle r^2 \rangle + K_i^{\text{MS}}  \frac{m_A - m_{A'}}{m_A m_{A'}} ,  
\label{IS}
\end{eqnarray}
where $\delta \langle r^2 \rangle = \langle r^2 \rangle_A - \langle r^2 \rangle_{A'}$ is the difference between the nuclear mean-square charge radii of the two isotopes \cite{King2013, Campbell2016}.
Higher-order effects and non-linear corrections to expression \ref{IS} are expected to be lower than 1$\%$ \cite{SELTZER1969}, and are thus neglected in our present study. The atomic part is factorized in the constants $F_i$ and $K_i^{\text{MS}}$, which are the FS and mass shift (MS) contributions to the LIS, respectively.
{\color{black}
The FS factor, $F_i = \frac{\langle \Psi_i | \sum_e F(r_e) | \Psi_i \rangle}{\langle \Psi_i | \Psi_i \rangle }$, for atomic level, $i$, described by the wave function, $|\Psi_i\rangle$, is calculated using the operator defined by 
\begin{eqnarray}
F(r_e) &=& - \frac{\delta V_{\text{nuc}}(r_N,r_e)}{\delta \langle r_N^2 \rangle} \;,
\end{eqnarray}
where $r_N$ is the nuclear radius ($\langle r_N^2 \rangle$ is the mean) and $r_e$ is the electronic coordinate.
The electrostatic potential due to the nuclear charge, $V_{\text{nuc}}(r_N,r_e)$, is evaluated by assuming a spherically-symmetric Fermi nuclear charge distribution defined by 
\begin{equation}
\rho_{\text{nuc}}(r_N)=\frac{\rho_{0}}{1+e^{(r_N-c)/a}} \; \; ,
\end{equation}
for the normalization factor, $\rho_0$. $c$ is the half-charge radius and $a$ is related to the skin thickness \cite{Hofstadter}.
The total MS constant is expressed as the sum of the normal MS (NMS), $K_i^{\text{NMS}}=\frac{\langle \Psi_i | \sum_e H_{NMS}(r_e) | \Psi_i \rangle}{\langle \Psi_i | \Psi_i \rangle }$, and specific MS (SMS), $K_i^{\text{SMS}}=\frac{\langle \Psi_i | \sum_{k,l\ge k} H_{SMS}(r_{kl}) | \Psi_i \rangle}{\langle \Psi_i | \Psi_i \rangle }$ for the inter-electronic distance, $r_{kl}=|\vec{r}_k - \vec{r}_l|$, between the electrons located at $r_k$ and $r_l$. These constants are obtained using the relativistic expressions of the operators given by \cite{Shabaev1994}
\begin{eqnarray}
H_{NMS}(r_i) =\vec{p}_i^{\; 2} - \frac{\alpha_e Z}{r_i} \vec{\alpha}_i^D \cdot \vec{p}_i - \frac{\alpha_e Z}{r_i} \left \{ (\vec{\alpha}_i^D \cdot \vec{C}_i^{(1)}) \vec{C}_i^{(1)} \right \} \cdot \vec{p}_i \; \; ,
\end{eqnarray}
and
\begin{eqnarray}
H_{SMS}(r_{ij}) = \vec{p}_i \cdot \vec{p}_j - \frac{\alpha_e Z}{r_i} \vec{\alpha}_i^D \cdot \vec{p}_j - \frac{\alpha_e Z}{r_i} \left \{ (\vec{\alpha}_i^D \cdot \vec{C}_i^{(1)}) \vec{C}_i^{(1)}\right \} \cdot \vec{p}_j \; \; .
\end{eqnarray}
In the above expressions, $\vec{p}$ is the momentum operator, $\alpha_e$ is the fine structure constant, $Z$ is the atomic number, $\vec{\alpha}^D$ is the Dirac matrix and $\vec{C}^{(1)}$ is the Racah operator of rank one.
It is worth noting here is that these expressions in the non-relativistic limit become $H_{NMS}(r_i)= \vec{p}_i^{\; 2}$ and $H_{SMS}(r_{ij}) = \vec{p}_i \cdot \vec{p}_j$. Since  $H_{SMS}$ is a two-body operator, evaluation of $K_i^{SMS}$ using the expectation value expression is computationally cumbersome.

\section{The relativistic coupled cluster theory and the analytic-response approach}

Traditionally, the finite-field (FF) approach is adopted through a suitable many-body method for the determination of IS factors, like the configuration-interaction (CI) approach, as they involve both the one-body and the two-body operators.
It is also observed that evaluation of expectation value of $\vec{p}^{\; 2}$ exhibits strong electron-correlation effects. This introduces difficulties in calculating using either the FF and expectation-value-evaluation (EVE) approaches, as the calculations do not converge with the inclusion of higher-order effects in the atomic wave functions \cite{Sahoo2019}.
In fact, this is also one of the reasons $\langle \vec{p}^{\; 2} \rangle$ is often approximated from the experimental energy in the heavy atomic systems following the Virial theorem \cite{Fock1930}.
As pointed out in Refs. \cite{Cubiss2018,Sahoo2010}, it is imperative to include both pair-correlation and core-polarization effects rigorously for accurate calculations of the IS.
The coupled cluster (CC) method incorporates both these effects to all orders.
Moreover, a truncated CC method, unlike a truncated CI method, is free from the size-extensivity and size-consistency problems appearing in many-body methods (e.g. see Ref. \cite{Bishop1991}).
In this work, we apply relativistic CC (RCC) theory to account for the relativistic effects in our calculations.}

\subsection{Basic aspects}

The atomic wave function of a state in an atomic system with a closed-shell configuration and with a valence orbital ($v$) can be expressed in the RCC theory as (e.g. see Refs. \cite{Sahoo2019,Sahoo2010,Sahoo2015} and therein)
\begin{eqnarray}
 |\Psi_v \rangle \equiv e^{\{T+S_v\}} |\Phi_v \rangle = e^T \{1+S_v\} |\Phi_v \rangle,
 \label{eqcc}
\end{eqnarray}
where $|\Phi_v \rangle= a_v^+ |\Phi_0\rangle$ with the Dirac-Hartree-Fock (DHF) wave function, $|\Phi_0 \rangle$, of the closed-core (in this work $[4d^{10}5s^2]$).
Here $T$ is the RCC excitation operator embodying electron-correlation effects from $|\Phi_0\rangle$ and the $S_v$ operator incorporates correlation of the electron from the valence orbital along with the core-valence interactions. 
Amplitudes of the RCC operators and energies are obtained using the following equations
\begin{eqnarray}
  \langle \Phi_0^L | (H e^T)_c | \Phi_0 \rangle = 0 \; ,
  \label{eqt0} 
\end{eqnarray} 
and
\begin{eqnarray}
 \langle \Phi_v^L | (He^T)_c  S_v | \Phi_v \rangle = E_v \langle \Phi_v^L | S_v | \Phi_v \rangle - \langle \Phi_v^L | (He^T)_c  | \Phi_v \rangle   , \ \ \ \
 \label{eqamp}
\end{eqnarray}
where $H$ is the atomic Hamiltonian and the subscript $c$ indicates connected terms.
The superscript, $L$, over the reference states indicates $L^{\text{th}}$-excited determinants with respect to the reference determinants appearing in the ket states. 
$E_0$ and $E_v$ are the exact energies of the states containing the closed-core (i.e. for the In$^+$ ion) and the closed-core with valence orbital, $v$, (i.e. for the In atom), respectively.
Both the $T$ and $S_v$ RCC operators are normal ordered with respect to $|\Phi_0 \rangle$.
For convenience we carry out all the calculations using normal-ordered operators, designated by subscript $N$.
The normal-ordered Hamiltonian is defined as $H_N=H-\langle \Phi_0 | H|\Phi_0 \rangle$, for the DHF energy, $E_{DHF}=\langle \Phi_0 | H|\Phi_0 \rangle$, using which the above amplitude solving equations for the RCC operator are given by
\begin{eqnarray}
  \langle \Phi_0^L | \bar{H}_N | \Phi_0 \rangle = 0 \; ,
  \label{eqt00} 
\end{eqnarray} 
and
\begin{eqnarray}
 \langle \Phi_v^L | \bar{H}_N  S_v  | \Phi_v \rangle = \Delta E_v \langle \Phi_v^L | S_v | \Phi_v \rangle  - \langle \Phi_v^L | \bar{H}_N | \Phi_v \rangle \; . \ \ \ \
 \label{eqamp0}
\end{eqnarray}
Here $\bar{H}_N=(H_N e^T)_c$, $\Delta E_0 =E_0 - E_{DHF}$ is the correlation energy of the closed core and $\Delta E_v = E_v - E_0$ is the electron affinity (EA) of the electron in the valence orbital, $v$.
We are interested in the EA values in this work, which are evaluated by 
\begin{eqnarray}
  \Delta E_v = \langle \Phi_v | \bar{H}_N  \{ 1 + S_v \} | \Phi_v \rangle  \; .
 \label{eqeng}
\end{eqnarray}
It is clear from the above that both Eqs. (\ref{eqamp0}) and (\ref{eqeng}) are correlated.
In our calculations we have considered Dirac-Coulomb-Breit (DCB) interactions in the atomic Hamiltonian, $H^a$. Further, we only consider all possible single- and double-excitation configurations in our RCC theory (RCCSD method). Excitation energy between two states are estimated from the difference of their EA values.

\subsection{The finite-field approach to isotope shifts}

Since all the relevant FS, NMS and SMS operators are scalar, they can be included with the atomic Hamiltonian to estimate their contributions to the energies.
On the other hand, by expressing the total Hamiltonian as $H=H^a + \lambda_v^O O$ with the atomic DCB Hamiltonian, $H^a$, and $O$, representing one of the FS, NMS or SMS operators for an arbitrary parameter, $\lambda_v^O$, it is possible to express the energy (here EA) in the FF approach as
\begin{eqnarray}
\Delta E_v= \Delta E_v^{(0)} + \lambda_v^O \Delta E_v^{(1)} + {\cal O}(\lambda_v^O)^{2} \; .
\end{eqnarray}
The superscripts (0), (1), and ${\cal O}(\lambda_v^O)^{2}$ denote the zeroth, first and higher-order  contributions respectively. It can be noted that the ${\cal O}(\lambda^O)^2$ contributions are not of our interest. It clearly follows that
\begin{eqnarray}
<O> \equiv \Delta E_v^{(1)} \simeq \left. \frac{\partial \Delta E_v}{\partial \lambda_v^O} \right|_{\lambda_v^O=0} .
\label{eqe1}
\end{eqnarray}
This obviously follows the Hellmann-Feynman (H-F) theorem \cite{Hellmann1935,Feynman1939}, but it has two major problems. First, the behaviors of FS, NMS and SMS operators are very different, the choice of $\lambda_v^O$ has to be distinct for estimating the FS, NMS and SMS constants reliably, and they can also be atomic state dependent.
Secondly, we assume ${\cal O}(\lambda^O)^2$ contributions are neglected in the FF approach based on the choice of the $\lambda_v^O$ value without removing them.
Usually the electron correlation effects contribute significantly to these quantities. Therefore, the IS constants inferred from the FF approach are subjected to large numerical uncertainty.
Nevertheless, we use $\lambda_v^O=1\times 10^{-6}$ to determine all the IS constants to perform the calculations in different states only for making comparative analysis of the results in our study.

\subsection{The expectation-value-evaluation approach}
{\color{black}
One can find several recent works that present high-precision results of many properties in atomic systems, e.g. hyperfine structure constants \cite{Sahoo2015,GarciaRuiz2018a}, by employing the RCC theory. These calculations are carried out using the EVE approach. Since the IS constants are the expectation values of the respective operators, we can evaluate them in the EVE approach using the RCC theory expression 
\begin{eqnarray}
< O >  \equiv \frac{\langle \Psi_v | O | \Psi_v \rangle}{\langle \Psi_v | \Psi_v \rangle} =\frac{\langle\Phi_v | \{1+S_v^{\dagger} \} e^{T\dagger } O_N e^T \{1+S_v\} | \Phi_v \rangle} {\langle\Phi_v | \{1+S_v^{\dagger} \} e^{T\dagger} e^T \{1+S_v\} | \Phi_v \rangle}
\label{eqcco}
\end{eqnarray}
by determining the wave functions using the Hamiltonian $H\equiv H^a$. The advantage of using this approach is that it is possible to analyse and observe the roles of various physical effects to the determination of the properties, whereas one can obtain only the final results in the FF approach without actually understanding the behavior of electron-correlation effects explicitly.
Evidently, this approach too has many shortcomings.
First, both the numerator and denominator of the above expression have non-terminating series. Secondly, the SMS operator is a two-body operator, so its  normal-ordered form will have two components in the calculations as (e.g. refer to \cite{Sahoo2010})
\begin{eqnarray}
O_N \equiv O_N^1 + O_N^2,
\end{eqnarray}
where superscipts 1 and 2 correspond to the effective one-body and two-body parts.
For the properties that are described by one-body operators, such as hyperfine structure constants, we have adopted an iterative procedure to account for contributions from the aforementioned non-terminating series in the numerator and denominator \cite{Sahoo2015}. However, it is impractical to apply a similar technique for the effective two-body terms, as it becomes unmanageable to compute contributions from the two-body components of the SMS operator using a diagrammatic procedure even in the RCCSD method approximation.
Thus, we estimate contributions by selecting only important diagrams representing the two-body components of the SMS operator based on the knowledge gained from our earlier studies (see discussions in Ref. \cite{Sahoo2010}). This may lead to large errors in the results. The third notable drawback of the EVE approach is, it does not satisfy the H-F theorem \cite{Bishop1991}.
This can be understood from the simple argument of Thouless \cite{Thouless1972}, that the form of Eq.~(\ref{eqcco}) does not follow the energy-evaluating expression given by Eq.~(\ref{eqeng}).

\subsection{The analytic-response approach}

The aforementioned problems of
(i) unwanted contributions from ${\cal O}(\lambda_v^O)^2$ in the FF approach,
(ii) the appearance of non-terminating series in the EVE approach,
(iii) the analysability of the roles of various physical effects in the determination of properties, and 
(iv) satisfying H-F theorem in the determination of the IS constants using the RCC theory, can all be circumvented by adopting the AR procedure as suggested by \cite{Monkhorst1977}.
}
The uniqueness of this approach is it uses features from both the FF and EVE procedures, in which Eq. (\ref{eqe1}) is directly obtained by perturbing the RCC operators due to $O$ as 
\begin{eqnarray}
T = T^{(0)} + \lambda_v^O T^{(1)} + {\cal O}(\lambda_v^O)^2 ,
\label{T_op_expan}
\end{eqnarray}
and 
\begin{eqnarray}
S_v = S_v^{(0)} + \lambda_v^O S_v^{(1)} + {\cal O}(\lambda_v^O)^2 ,
\label{S_op_expan}
\end{eqnarray}
where $T$ and $S_v$ are the RCC operators for the total Hamiltonian, $H=H^a + \lambda_v^O O$, and superscripts $(0)$ and $(1)$ indicate the unperturbed and the first-order perturbed corrections due to $O$, respectively.
Substituting the above expanded form of the operators into Eqs. (\ref{eqt00}) and (\ref{eqamp0}), and then equating the zeroth-order and first-order terms in $\lambda^O$ gives the equations for the unperturbed and perturbed RCC operators, respectively. {\color{black}
Similarly, the first-order terms from the expansion in Eq. (\ref{eqeng}) will correspond to the expectation values of the operator $O$. Thus, using the normal-ordered form of the operators, we can get
\begin{eqnarray}
  \langle \Phi_0^L | \bar{H}^a_N T^{(1)} | \Phi_0 \rangle &=& - \langle \Phi_0^L | \bar{O}_N| \Phi_0 \rangle , 
  \label{eqt1} \\
 \langle \Phi_v^L | (\bar{H}^a_N-\Delta E_v^{(0)}) S_v^{(1)} |\Phi_v \rangle &=& \Delta E_v^{(1)} \langle \Phi_v^L | S_v^{(0)} | \Phi_v \rangle \nonumber , \\ && - \langle \Phi_v^L | ( \bar{H}^a_N T^{(1)} + \bar{O}_N) \{ 1+ S_v^{(0)} \} | \Phi_v \rangle  , 
 \label{eqamp1} 
\end{eqnarray}
and 
\begin{eqnarray}
\Delta E_v^{(1)} =  \langle \Phi_v | \bar{H}_N S_v^{(1)} + (\bar{H}_N T^{(1)}+ \bar{O}_N) \{ 1+ S_v^{(0)} \}  | \Phi_v \rangle .
\label{eqe11}
\end{eqnarray}
Here, $\bar{O}_N=(O_N e^{T^{(0)}})_c$, and the superscripts $(0)$ and $(1)$ in the energies indicate the zeroth and first-order contributions, respectively. 
The AR equations have the advantages that were mentioned above. 
It can be noted that the lowest-order contributions (DHF results) in the EVE and AR approaches are the same, while they are different in the FF procedure.
Again, the above equations are modified appropriately for the evaluation of the SMS constants as
\begin{eqnarray}
  \langle \Phi_0^L | \bar{H}^a_N T^{(1)} | \Phi_0 \rangle &=& - \langle \Phi_0^L | \bar{O}_N^1 + \bar{O}_N^2 | \Phi_0 \rangle , 
  \label{eqt21} \\
 \langle \Phi_v^L | (\bar{H}^a_N-\Delta E_v^{(0)}) S_v^{(1)} |\Phi_v \rangle &=& \Delta E_v^{(1)} \langle \Phi_v^L | S_v^{(0)} | \Phi_v \rangle  \nonumber \\  &-& \langle \Phi_v^L | ( \bar{H}^a_N T^{(1)} + \bar{O}_N^1 + \bar{O}_N^2) \{ 1+ S_v^{(0)} \} | \Phi_v \rangle   ,
 \label{eqamp21} 
\end{eqnarray}
and 
\begin{eqnarray}
\Delta E_v^{(1)} =  \langle \Phi_v | \bar{H}_N S_v^{(1)} + (\bar{H}_N T^{(1)}+ \bar{O}_N^1 +\bar{O}_N^2 ) \{ 1+ S_v^{(0)} \}  | \Phi_v \rangle ,
\label{eqe211}
\end{eqnarray}
due to the two-body nature of the SMS operator.
The AR approach also involves a slight computational challenge compared with the FF and EVE approaches as it requires storing matrix elements of the additional one-body and two-body operators than the atomic Hamiltonian. 

Ideally, if results from all the three, FF, EVE and AR, approaches agree with each other then the results can be assumed as very reliable.
However, it is difficult to achieve good agreement between the results from all these procedures in heavy atomic systems using approximated many-body methods and due to large numerical uncertainties associated with the implementation of the EVE and FF approaches.
Nonetheless, the results obtained using the AR approach at the given level of approximation in the many-body theory should be treated as more valid owing to the aforementioned merits of this procedure. 
}

\Table{\label{fields_table} Comparison of FS, NMS and SMS factors of the six states in indium from the FF, EVE and AR approaches obtained using the RCCSD method.}
\br
Method &  5P$_{1/2}$ & 5P$_{3/2}$ & 6S$_{1/2}$ & 7S$_{1/2}$ & 8S$_{1/2}$ & 9S$_{1/2}$ \\
\br
\multicolumn{7}{c}{$F$ (GHz/fm$^2$)} \\
\br
 FF & 1.544 & 1.491 & -0.437 & -0.155 & -0.069 & -0.033 \\
 EVE & 1.275 & 1.299 & -0.408 & -0.135 & -0.061 & -0.033 \\
 AR & 1.435(6) & 1.442(6) & -0.383(1) & -0.1281(5) & -0.0559(25) & -0.0307(5) \\
\br
\multicolumn{7}{c}{$K_{\text{NMS}}$ (GHz.u) } \\
\br
 FF & 749 & 711 & 364 & 170 & 98 & 63 \\
 EVE & 1340 & 375 & 458 & 201 & 113 & 71 \\
 AR & 774(41) & 734(37) & 340(5) & 163(2) & 96(1) & 61.7(5) \\            
 Experiment$^\dagger$ & 768 & 731 & 367 & 171 & 99 & 65 \\
\br
\multicolumn{7}{c}{$K_{\text{SMS}}$ (GHz.u)} \\
\br
 FF & -470 & -403 & 119 & 38 & 17 & 9 \\
 EVE & -1048 & -899 & 136 & 42 & 18 & 10 \\
 AR & -638(71) & -533(69) & 94(26) & 29(8) & 13(4) & 8.6(5) \\
 Experiment$^*$ & -536(122) & -507(111) & 169(51) & 55(42) & 24(80) & -13(66) \\
\br
\multicolumn{7}{c}{$\text{LIS}^{113, 115}$ (MHz)} \\
\br
Experiment & 277(10) & 272(6)\cite{Menges1985} & 17(6) & 12(6) & 9(12) & 2(10)\\
\br
\end{tabular}
\\ 
\item[]   $\dagger$ Level energies from \cite{nist} were used.
\item[]   \text{*} To determine $K_{\text{SMS}}$ from Eq. \ref{IS}, the measured differential ISs, $\delta E^{113, 115}$, were combined with FS factors from the AR approach and $\delta \left\langle r^2 \right\rangle _{\mu}^{113, 115}$~=~0.157(11) fm$^2$ \cite{Fricke}
\end{indented}
\end{table}

\section{Isotope shift measurements}

The results of the calculations have been combined with complementary measurements to perform a comprehensive theoretical and experimental study of the FS and SMS constants of the indium atom.
Further, they are used to provide accurate nuclear charge-radii of $^{104-127}$In.
As indium has only two naturally occurring isotopes ($^{113,115}$In), exotic isotopes were produced at the on-line isotope-separator facility ISOLDE at CERN.
To produce the neutron-rich indium isotopes, $^{115-127}$In, a beam of 1.4-GeV protons impinged onto the neutron converter of a thick UC$_x$ target.
The converter suppressed nearby caesium mass contamination and increased utilizable neutron-rich indium yields \cite{Dillmann2002}.
The neutron-deficient indium isotopes, $^{104-115}$In, were produced by impinging the protons directly onto a thick LaC$_2$ target \cite{Koster2002}.
The indium isotopes diffused through the target material and their ionization was enhanced by the use of the resonant ionization ion source RILIS \cite{Rothe2011}.
The produced \cite{Rothe2011} indium ions were then accelerated to 40~keV, mass separated, and injected into a gas-filled linear Paul trap (ISCOOL) \cite{Mane2009, Franberg2008}.
Ion bunches of 2~$\mu$s temporal width, were then re-accelerated to 40 keV and deflected into the CRIS beamline \cite{Flanagan13, Cocolios2016}.
The indium ions were then neutralised with a sodium-filled vapor cell, with an efficiency of up to 60\% and predicted relative atomic populations of 57\% and 37\% respectively for the 5p~$^2$P$_{3/2}$ metastable state and 5p~$^2$P$_{1/2}$ ground state \cite{Vernon2019}.
The remaining ion fraction was removed by electrostatic deflectors, and the neutralized atom bunch was collinearly overlapped with two pulsed lasers, one for excitation and another for non-resonant ionization.
The atoms were resonantly excited using two different UV transitions in separate measurements.
The first using 246.8-nm laser light for the 5p~$^2$P$_{3/2}$ $\rightarrow$ 9s~$^2$S$_{1/2}$ atomic transition.
The second using 246.0-nm laser light for the 5p~$^2$P$_{1/2}$ $\rightarrow$ 8s~$^2$S$_{1/2}$ atomic transition.
The resonant laser light was produced by frequency tripling the light from an injection-locked Ti:Sapphire laser system \cite{Sonnenschein2017}. 
This laser was seeded using a narrow-band M Squared SolsTiS continuous-wave Ti:Sapphire laser, and pumped using a LEE LDP-100MQ Nd:YAG laser, producing pulsed narrow-band 740(738)-nm laser light at 1 kHz. 
This light was then frequency tripled to 246.8(246.0)-nm light by the use of two non-linear BiB$_3$O$_6$ crystals \cite{Bass1962}, 3~mW of laser light was used to saturate both transitions.
The excited atoms were then ionized by a non-resonant 532-nm step, 
The frequency of the resonant first step was scanned and the resulting ions were deflected onto a detector, producing the hyperfine spectra from which the IS were obtained.
The determined IS values are displayed in Table~\ref{tab:IS_delrs}.\\

\section{Comparison with experiment and evaluation of nuclear mean-squared charge radii}
\subsection{King plot analysis}
Since the changes in the mean-square charge radii are independent of the atomic transitions, the nuclear dependence can be removed by comparing the IS of two atomic transitions.
A combination of the IS using Eq. (\ref{IS}), for two atomic transitions, $i$ and $j$, can be expressed as

\begin{equation}
\mu_{A,A'}\delta E^{A,A'}_j=\frac{F_j}{F_i}\mu_{A,A'}\delta E^{A,A'}_i+ M_j-\frac{F_j}{F_i}M_i,
\label{modIS}
\end{equation}

with $\mu_{A, A'} = \frac{m_A m_{A'}}{m_A - m_{A'}}$.
Hence, in a `King' plot \cite{King2013} of $\mu_{A,A'}\delta E^{A,A'}_j$ versus $\mu_{A, A'} \delta E^{A, A'}_i$, the gradient provides the FS ratio, $F_j/F_i$, between two transitions, and the MS differences can be extracted from its intercept.

\begin{figure}
\centering
\includegraphics[width=11cm]{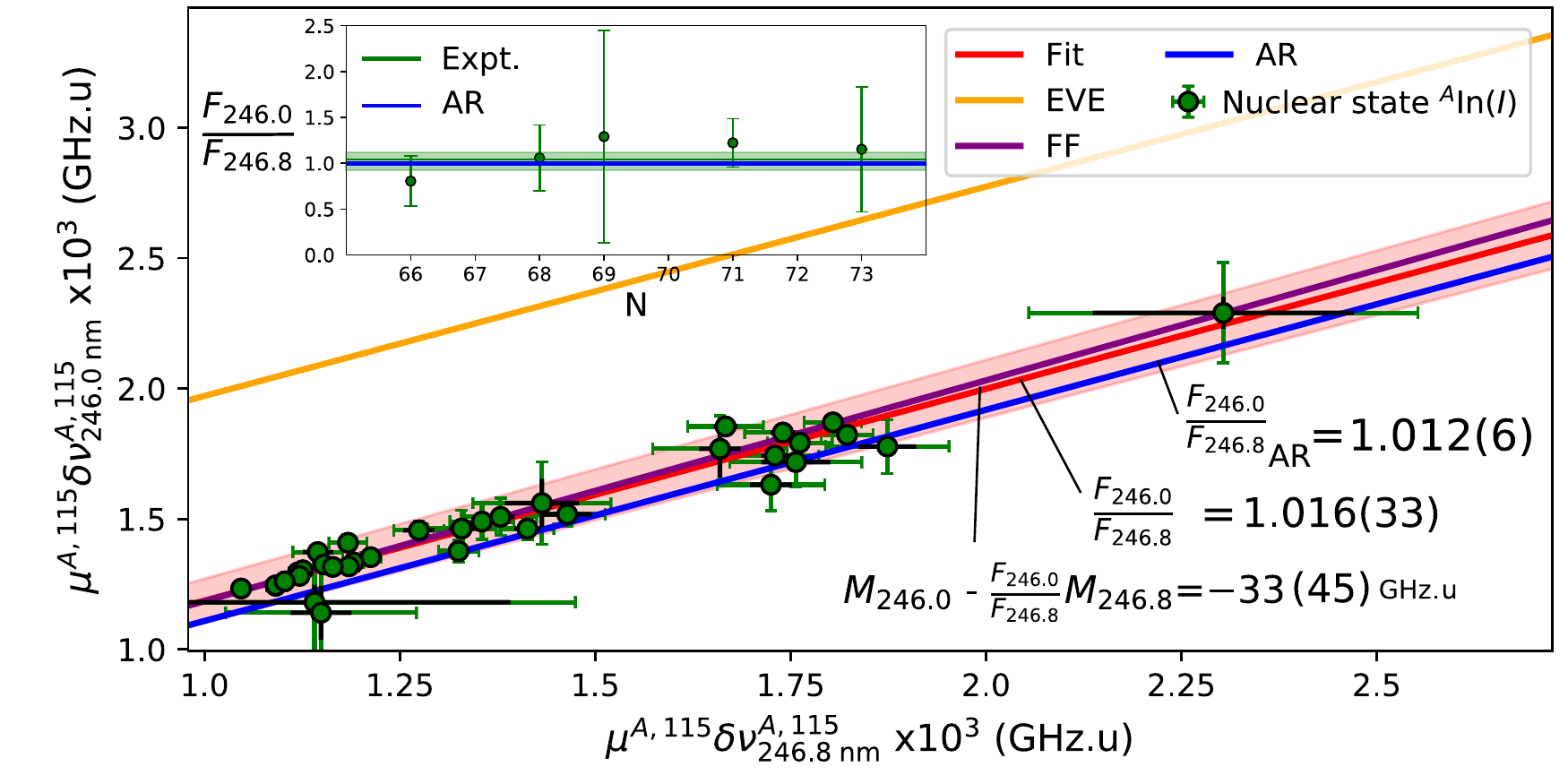}
\includegraphics[width=11cm]{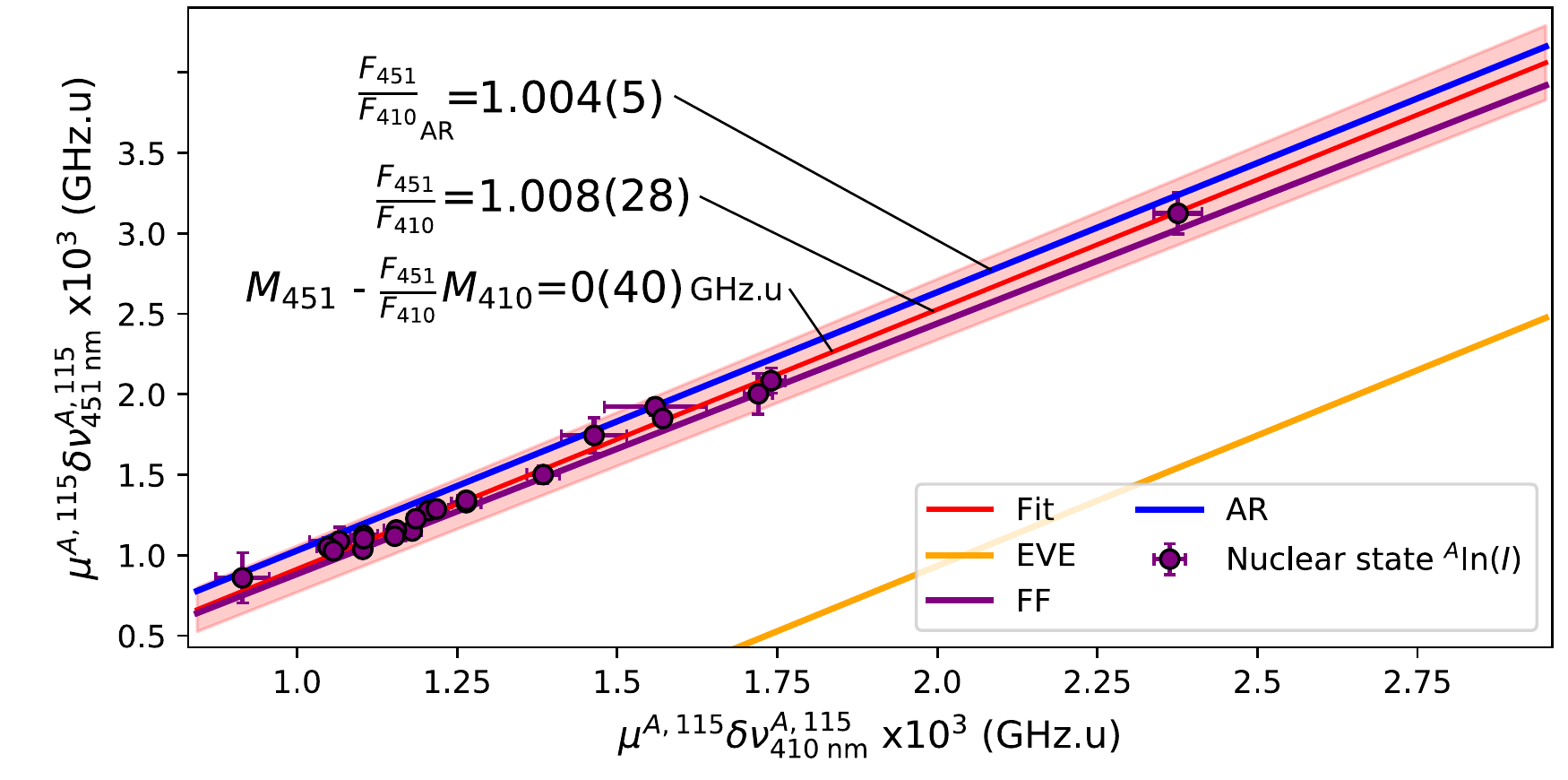}
\caption{\label{king_plots}
King plots of the 246.0-nm and 246.8-nm and the 410.2-nm and 451.1-nm transitions.
Inset: the ratio of isomer shift values allowed mass-shift-independent determination of $\frac{F_{246.0}}{F_{246.8}}=1.04(9)$.
Theoretical values are indicated by $\frac{F_{246.0}}{F_{246.8}}_{AR}$.
The shaded area indicates the uncertainty of the fits.
Error bars include statistical and systematic uncertainties (indicated by the black part of the error bar)}
\end{figure}

The King plot obtained for the transitions measured in this work (246.8-nm (5p~$^2$P$_{3/2}$ $\rightarrow$ 9s~$^2$S$_{1/2}$) and 246.0-nm (5p~$^2$P$_{1/2}$ $\rightarrow$ 8s~$^2$S$_{1/2}$)), and previous measurements in the 410.2-nm (5p~$^2$P$_{1/2}$ $\rightarrow$ 6s~$^2$S$_{1/2}$) and 451.1-nm (5p~$^2$P$_{3/2}$ $\rightarrow$ 6s~$^2$S$_{1/2}$) transitions \cite{Eberz1987a} are shown in Fig.~\ref{king_plots}.
The calculations and experimental data agree within 1$\sigma$, using the AR or FF approaches.

\subsection{Isomer shifts}
The availability of several isomeric nuclear states in the indium isotope chain allows a further test of the theoretical calculations.
For isomeric states, the factor $\frac{m_A - m_{A'}}{m_A m_{A'}}$ tends to 0, and the Eq. (\ref{modIS}) can be approximated as $\frac{\delta E^{m}_i}{\delta E^{m}_j} = \frac{F_i}{F_j}$.
This assumption corresponds to an uncertainty of up to 0.02~MHz for the excitation energies of the isomers in this work (\textless400 keV).
Therefore, isomer-shift measurements provide a test of the FS factors and are less sensitive to systematic uncertainties present in the King plot analysis. 
Previous measurements have not reported values for isomer shifts in the indium atom as they are relatively small and require particularly high precision \cite{Eberz1987a}.
The new measurements reported here allowed the extraction of isomer shifts for the 246.8-nm (5p~$^2$P$_{3/2}$ $\rightarrow$ 9s~$^2$S$_{1/2}$) and 246.0-nm (5p~$^2$P$_{1/2}$ $\rightarrow$ 8s~$^2$S$_{1/2}$) transitions.
The extracted FS ratios from the measured isomer shifts are shown in the inset of Fig.~\ref{king_plots}. 
This ratio agrees with the value obtained from the King plots, and is within 1$\sigma$ of the presented theoretical calculations.

\subsection{Experimental level specific mass shift}

Calculations of SMS are notably challenging. 
To the authors' knowledge, they have not yet been reported for the indium atom.
Moreover, a reliable experimental test is also difficult as optical measurements provide the difference of SMS between two states and their individual contribution cannot be separated.
Yet calculations of the atomic FS and MS factors are typically performed for individual atomic-energy levels, with the difference between two states used to determine the atomic factors for a transition used to measure an isotope shift.
In this work the individual atomic-level isotope shift (LIS) values were determined by combining the IS measurements with measurement of transitions to high-lying atomic states in indium \cite{Menges1985}.
As the contribution to the IS of a transition from an atomic state decreases with the principle quantum number of the state, in measurements to high-lying Rydberg states the IS contribution from the upper state becomes negligible \cite{Niemax1980}.
This allowed the LIS to be determined for each state, and then the specific-mass-shift contribution to the individual state, $l$, could be evaluated for comparison to the calculations.
The new measurements of this work provide access to the 8S$_{1/2}$ and 9S$_{1/2}$ states.
For example, using the 5p~$^2$P$_{3/2}$~$\rightarrow$~5s$^2$~np~$^2$P$_{1/2, 3/2}$ transition (27$\leq$n$\leq$35) IS measured for $^{113,115}$In \cite{Menges1985}, a LIS of the 5p~$^2$P$_{3/2}$ state of $\text{LIS}_{\text{P3/2}}^{113, 115}$=272(6)~MHz was reported. 
Using the IS value measured with the 5p~$^2$P$_{3/2}$ $\rightarrow$ 6s~$^2$S$_{1/2}$ transition of $\text{LIS}_{\text{P3/2-6s}}^{113, 115}$=255.4(5)~MHz \cite{Zaal1978}, gives a LIS of $\text{LIS}_{\text{6s}}^{113, 115}$~=~17(6)~MHz. 
This LIS value can in turn be used to determine the LIS of the 5p~$^2$P$_{1/2}$ state from the 5p~$^2$P$_{1/2}$ $\rightarrow$ 6s~$^2$S$_{1/2}$ transition \cite{GarciaRuiz2018a}, giving $\text{LIS}_{\text{5P1/2}}^{113, 115}$~=~277(10)~MHz.
All of the $\text{LIS}$ values determined from the new measurements of this work and from literature (6S$_{1/2}$ and 7S$_{1/2}$ states \cite{GarciaRuiz2018a, Zaal1978, Eliel1981}) are presented in Table~\ref{fields_table}.
The LIS value, $\text{LIS}^{113, 115}_l$, of a state, $l$, is the sum of the field-shift (volume isotope shift) and mass-shift contributions given by

\begin{align}
\begin{aligned}
    \text{LIS}^{113, 115}_l &=  F_l \delta \langle r^2 \rangle^{113,115}\\
      &+ (K_l^{\text{NMS}}+K_l^{\text{SMS}})\frac{m_{113}-m_{115}}{m_{113} m_{115}}
\end{aligned}
\label{ISsupp}
\end{align}

Using the calculated state FS atomic factors and relativistic normal mass shift factors, $K_l^{\text{NMS}}$, given in Table~\ref{fields_table}, and the literature value of $\delta \left\langle r^2 \right\rangle _{\mu}^{113, 115}$~=~0.157(11)~fm$^2$ \cite{Fricke} allowed evaluation of the SMS factors for individual states, $K_{\text{SMS}}^{\text{Exp}}$.
The experimental results and theoretical calculations are shown in Table~\ref{fields_table}. 
The new calculations presented here, adopting the AR approach, agree within 1$\sigma$ of the experimental values, in addition to the values from the FF approach.
In contrast, the EVE results present large discrepancies.

\subsection{Comparison with nuclear mean-squared charge radii}

\begin{figure}[t]
\centering
\includegraphics[width=12cm]{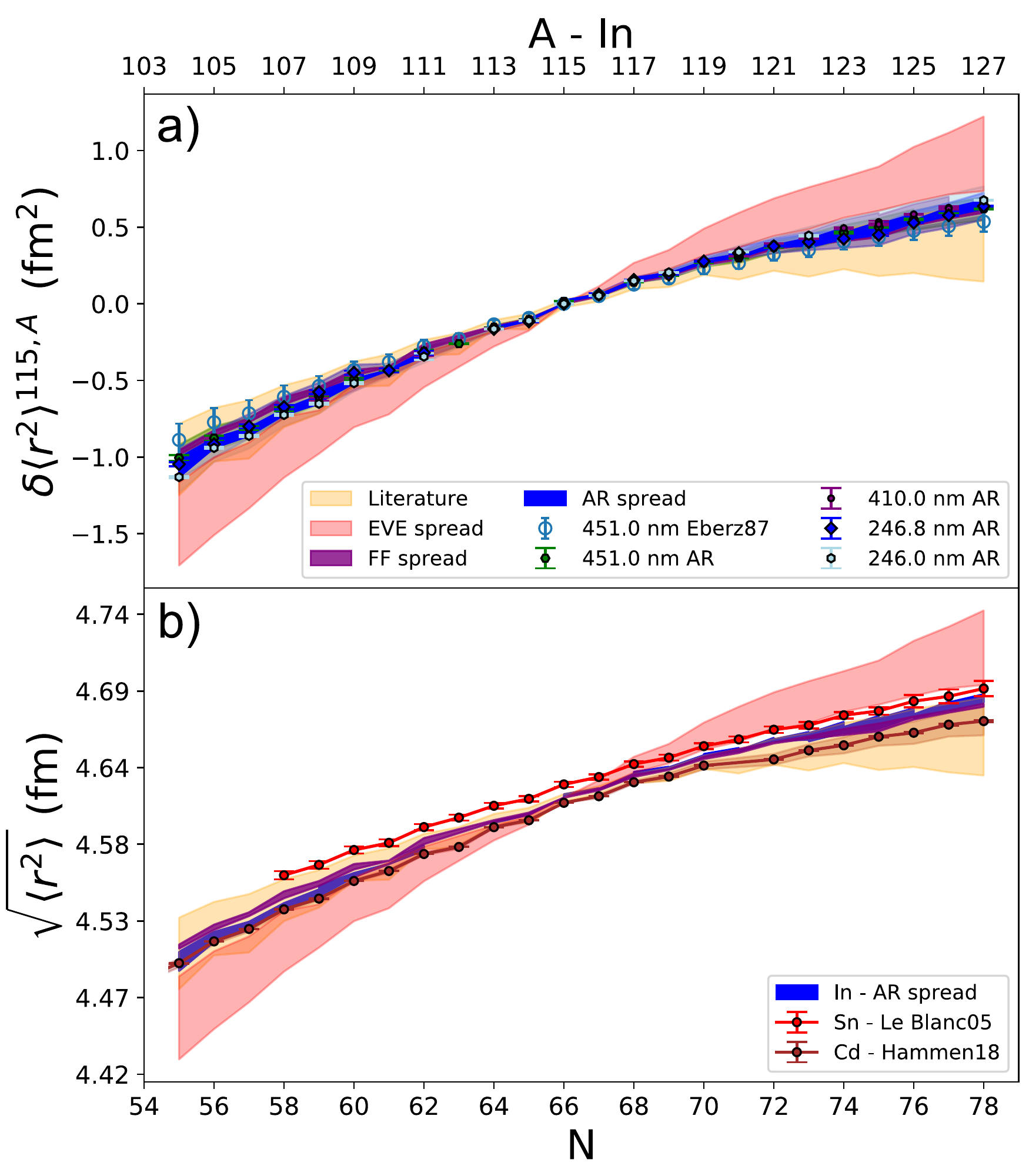}
\caption{\textbf{a)}\label{fig:col_rad_def} $\delta \left\langle r^2 \right\rangle^{115, A}$ values for the $^{104-127}$In isotopes extracted from the IS measurements of four optical transitions and using the calculated FS and MS factors. 
The spread in values from each approach is indicated by the colored areas.
The shaded area `Literature' indicates the uncertainty from literature FS and MS factors \cite{Fricke, Eberz1987a}. 
\textbf{b)} \label{fig:col_rad_abs} $\sqrt{\left\langle r^2 \right\rangle^{A}}$ compared to Sn ($Z$ = 50) \cite{LeBlanc2005} and Cd ($Z$ = 48) \cite{Hammen18} isotopes.}
\end{figure}

Combining the IS measurements and the calculated FS and MS constants in Eq.~(\ref{IS}), a value of $\delta \left\langle r^2 \right\rangle^{113, 115}$~=~0.163(4)~fm$^2$ is obtained for the root-mean-square charge radii difference between the stable isotopes $^{113,115}$In, in good agreement with the muonic atom result of $\delta \left\langle r^2 \right\rangle _{\mu}^{113, 115}$~=~0.157(11) fm$^2$ \cite{Fricke}. The nuclear charge radii of the exotic indium isotopes were extracted from the measured IS and the calculated FS and SMS constants from the AR approach.
The extracted $\delta \left\langle r^2 \right\rangle^{115, A}$ values are given in Table~\ref{tab:IS_delrs} and are plotted in Fig.~\ref{fig:col_rad_def}.
The reported uncertainties of the calculated atomic factors using the AR approach were evaluated from a perturbative estimation of the neglected triples contribution.
The atomic masses used were taken from \cite{Wang2017}.
The values obtained from the FF and EVE approaches are also shown in Fig.~\ref{fig:col_rad_def} for comparison. 

\Table{\label{tab:IS_delrs} IS measured with the 246.0-nm (5p $^2$P$_{1/2}$ $\rightarrow$ 8s $^2$S$_{1/2}$) and 246.8-nm ( 5p $^2$P$_{3/2}$ $\rightarrow$ 9s $^2$S$_{1/2}$ ) transitions, and $\delta \left\langle r^2 \right\rangle^{115, A}$ values extracted using the AR approach. }
\br
A & I & \multicolumn{2}{c}{$IS^{115, A}$ (MHz)} & \multicolumn{2}{c}{$\delta \left\langle r^2 \right\rangle^{115, A}$ (fm$^2$)} \\
& & 246.0 nm & 246.8 nm & 246.0 nm & 246.8 nm \\
\br
104 & ($5^+$) & -1805(10) & -1753(20) & -1.19(5) & -1.11(5) \\
105 & $\frac{9}{2}^+$ & -1510(10) & -1540(20) & -1.00(5) & -0.97(5) \\
106 & $7^+$ & -1381(10) & -1362(20) & -0.91(4) & -0.86(4) \\
107 & $\frac{9}{2}^+$ & -1166(10) & -1178(20) & -0.77(4) & -0.74(4) \\
108 & $2^+$ & -1033(10) & -978(20) & -0.68(3) & -0.61(3) \\
108 & $7^+$ & -1046(10) & -1011(20) & -0.69(3) & -0.64(3) \\
109 & $\frac{9}{2}^+$ & -835(10) & -855(20) & -0.55(3) & -0.54(3) \\
110 & $7^+$ &  & -729(20) &  & -0.46(2) \\
111 & $\frac{9}{2}^+$ & -555(30) & -542(20) & -0.37(3) & -0.34(2) \\
113 & $\frac{9}{2}^+$ & -265(5) & -278(5) & -0.175(9) & -0.175(9) \\
114 & $5^+$ & -175(5) & -171(10) & -0.116(5) & -0.109(8) \\
115 & $\frac{9}{2}^+$ & 0 & 0 & 0 & 0 \\
115 & $\frac{1}{2}^-$ & 26(8) & 33(5) & 0.018(5) & 0.022(3) \\
116 & $5^+$ & 89(5) & 99(20) & 0.058(5) & 0.06(1) \\
116 & $8^-$ & 86(8) & 99(2) & 0.056(7) & 0.061(4) \\
117 & $\frac{9}{2}^+$ & 243(5) & 265(3) & 0.160(9) & 0.167(8) \\
117 & $\frac{1}{2}^-$ & 261(6) & 282(4) & 0.173(9) & 0.179(8) \\
118 & $5^+$ & 330(5) & 329(2) & 0.22(1) & 0.20(1) \\
118 & $8^-$ & 324(5) & 324(3) & 0.21(1) & 0.20(1) \\
119 & $\frac{9}{2}^+$ &  & 475(3) &  & 0.30(2) \\
119 & $\frac{1}{2}^-$ &  & 488(4) &  & 0.30(2) \\
120 & $(5)^+$ & 531(5) & 556(5) & 0.35(2) & 0.35(2) \\
120 & $(8^-)$ & 500(5) & 530(2) & 0.33(2) & 0.33(2) \\
121 & $\frac{9}{2}^+$ &  & 654(2) &  & 0.41(2) \\
121 & $\frac{1}{2}^-$ &  & 661(3) &  & 0.41(2) \\
122 & $5^+$ & 704(5) & 674(5) & 0.46(3) & 0.41(3) \\
122 & $8^-$ & 687(5) & 658(8) & 0.45(3) & 0.40(3) \\
123 & $\frac{9}{2}^+$ &  & 756(3) &  & 0.46(3) \\
123 & $\frac{1}{2}^-$ &  & 751(2) &  & 0.46(3) \\
124 & $(3)^+$ &  & 809(10) &  & 0.49(3) \\
124 & $(8^-)$ &  & 810(3) &  & 0.49(3) \\
125 & $\frac{9}{2}^+$ &  & 941(4) &  & 0.58(4) \\
125 & $\frac{1}{2}^-$ &  & 926(5) &  & 0.57(4) \\
126 & $3^+$ &  & 1026(3) &  & 0.63(4) \\
126 & $(8^-)$ &  & 1019(5) &  & 0.62(4) \\
127 & $\frac{9}{2}^+$ & 1115(5) & 1129(4) & 0.73(5) & 0.69(4) \\
\br
\end{tabular}
\end{indented}
\end{table}

Remarkably, the extracted $\delta \left\langle r^2 \right\rangle$ values agree for all four optical transitions, which gives confidence in the accuracy of the calculations.
The absolute charge radii, $\sqrt{\left\langle r^2 \right\rangle^{A}}$, using the reference isotope $^{115}$In (4.615 fm \cite{Fricke}), are compared to its isobaric neighbours Sn ($Z$~=~50) \cite{LeBlanc2005} and Cd ($Z$~=~48) \cite{Hammen18} in Fig. \ref{fig:col_rad_abs}.
The effect of inaccurate calculation of the MS factors using the EVE approach is seen to be significant, causing a large discrepancy between the values extracted from the four transitions.
Previously, literature values \cite{Eberz1987a, Fricke} were normalized to the neighboring tin and cadmium isotopes and the $\delta \left\langle r^2 \right\rangle _{\mu}^{113, 115}$ value.
This introduces large uncertainties (yellow area in Fig. \ref{fig:col_rad_def}), and prevents an independent comparison of the nuclear charge radii with neighbouring elements.
Our theoretical calculations have therefore enabled the first independent comparison of absolute charge radii for an odd-proton system around the $Z$~=~50 nuclear closed shell to be made.

\section{Conclusion}
In conclusion, we present in this work a new theoretical method to perform accurate calculations of FS and MS constants in atomic systems.
These constants are critical to separate electronic and nuclear structure effects in the interpretation of IS measurements for fundamental and nuclear-physics research. 
This new theoretical method uses an analytic-energy-derivative approach in the RCC framework, and solves fundamental problems related to the evaluation of operators, which have been present in previous atomic physics calculations.
Precise IS measurements in the indium atom were used as an exhaustive experimental test for these theoretical developments. A good agreement was found with all available experimental data.
The existence of several isomers and the access to high-lying states in the indium atom allow the separation of FS from MS, providing a stringent test for the calculations.
Our calculations of the atomic physics factors are essential to extract nuclear charge radii values from isotope shifts measurements of exotic indium isotopes \cite{Ruiz2017a}.
These results can be extended to different elements across the nuclear chart.
This is especially important for odd-proton nuclei, which rely on atomic theory to extract charge radii from laser-spectroscopy measurements.
Our theoretical developments will help to provide a deeper insight in the evolution of the nuclear charge radius for different numbers of protons and neutrons, which is of great importance for our understanding of nuclear structure  \cite{Hammen18, Morris2018, garciaruiz16,ekstrom15} and nuclear matter \cite{Brown17}.

\section{Acknowledgments}
This work was supported by ERC Consolidator Grant No.648381 (FNPMLS); 
STFC grants ST/L005794/1, ST/L005786/1, ST/P004423/1 and Ernest Rutherford Grant No. ST/L002868/1; 
GOA 15/010 from KU Leuven, BriX Research Program No. P7/12; the FWO-Vlaanderen (Belgium); 
the European Unions Grant Agreement 654002 (ENSAR2); 
National Key R\&D Program of China (Contract No:2018YFA0404403), the National Natural Science Foundation of China (No:11875073) 
and we acknowledge the financial aid of the Ed Schneiderman Fund at New York University.
B. K. S. acknowledges use of the Vikram-100 HPC cluster of Physical Research Laboratory, Ahmedabad. 
We would also like to thank the ISOLDE technical group for their support and assistance, the University of Jyv\"askyl\"a for the use of the injection-locked cavity, and the Physikalisch-Technische Bundesanstalt (PTB) for the use of their voltage divider.

\section*{References}

\bibliographystyle{unsrt}
\bibliography{apssamp.bib}

\end{document}